\documentclass[aps,twocolumn,showpacs,preprintnumbers,amsmath,amssymb,superscriptaddress]{revtex4}

\usepackage{graphicx}
\usepackage{dcolumn}
\usepackage{bm}
\usepackage{epsfig}

\begin{document}

\title{$\alpha$-induced reactions on $^{115}$In: cross section measurements and
statistical model analysis}

\author{G.\,G.\,Kiss}%
 \email{ggkiss@atomki.mta.hu}
\affiliation{%
Institute for Nuclear Research (MTA Atomki), H-4001 Debrecen, POB.51., Hungary}%
\author{T.\,Sz\"ucs}%
\affiliation{%
Institute for Nuclear Research (MTA Atomki), H-4001 Debrecen, POB.51., Hungary}%
\author{P.\,Mohr}%
\affiliation{%
Institute for Nuclear Research (MTA Atomki), H-4001 Debrecen, POB.51., Hungary}%
\affiliation{%
Diakonie-Klinikum, D-74523 Schw\"abisch Hall, Germany}%
\author{Zs.\,T\"or\"ok}%
\affiliation{%
Institute for Nuclear Research (MTA Atomki), H-4001 Debrecen, POB.51., Hungary}%
\author{R.\,Husz\'ank}%
\affiliation{%
Institute for Nuclear Research (MTA Atomki), H-4001 Debrecen, POB.51., Hungary}%
\author{Gy.\,Gy\"urky}%
\affiliation{%
Institute for Nuclear Research (MTA Atomki), H-4001 Debrecen, POB.51., Hungary}%
\author{Zs.\,F\"ul\"op}%
\affiliation{%
Institute for Nuclear Research (MTA Atomki), H-4001 Debrecen, POB.51., Hungary}%

\date{\today}

\begin{abstract}
\begin{description}
\item[Background] Alpha-nucleus optical potentials are basic ingredients of
  statistical model calculations used in nucleosynthesis simulations. While
  the nucleon+nucleus optical potential is fairly well known, for the
  $\alpha$+nucleus optical potential several different parameter sets exist
  and large deviations, reaching sometimes even an order of magnitude, are
  found between the cross section predictions calculated using different
  parameter sets. 

\item[Purpose] A measurement of the radiative $\alpha$-capture and the
  $\alpha$-induced reaction cross sections on the nucleus $^{115}$In at low
  energies allows a stringent test of statistical model predictions. Since
  experimental data is scarce in this mass region, this measurement can be an
  important input to test the global applicability of $\alpha$+nucleus optical
  model potentials and further ingredients of the statistical model.
  
\item[Methods] The reaction cross sections were measured by means of the
  activation method. The produced activities were determined by off-line
  detection of the $\gamma$-rays and characteristic x-rays emitted during the
  electron capture decay of the produced Sb isotopes. The
  $^{115}$In($\alpha,\gamma$)$^{119}$Sb and $^{115}$In($\alpha$,n)$^{118}$Sb$^m$
  reaction cross sections were measured between $E_\mathrm{c.m.}$ = 8.83 MeV -
  15.58 MeV, and the $^{115}$In($\alpha$,n)$^{118}$Sb$^{g}$ reaction was
  studied between $E_\mathrm{c.m.}$ = 11.10 MeV - 15.58 MeV. The theoretical
  analysis was performed within the statistical model.
   
\item[Results] The simultaneous measurement of the ($\alpha,\gamma$) and
  ($\alpha$,n) cross sections allowed to determine a best-fit combination of
  all parameters for the statistical model. The $\alpha$+nucleus optical
  potential is identified as most important input for the statistical
  model. The best fit is obtained for the new Atomki-V1 potential, and good
  reproduction of the experimental data is also achieved for the first version
  of the Demetriou potentials and the simple McFadden/Satchler
  potential. The nucleon optical potential, the $\gamma$-ray strength
  function, and the level density parametrization are also constrained by the
  data although there is no unique best-fit combination. 

\item[Conclusions] The best-fit calculations allow to extrapolate the
  low-energy ($\alpha$,$\gamma$) cross section of $^{115}$In to the
  astrophysical Gamow window with reasonable uncertainties. However, still further
  improvements of the $\alpha$-nucleus potential are required for a global
  description of elastic ($\alpha$,$\alpha$) scattering and $\alpha$-induced
  reactions in a wide range of masses and energies.

\end{description}
\end{abstract}


\maketitle

\section{Introduction}
\label{sec:int}

\subsection{Nucleosynthesis simulations and optical potentials}
\label{sec:nuc}

Isotopes heavier than iron are synthesized by several astrophysical processes
at various sites. The bulk of these isotopes are formed by two neutron capture
processes: slow neutron capture (s-process) \cite{Kap11} and rapid neutron
capture process (r-process) \cite{Thie11}. However, other processes --- such
as i-process \cite{Cow77, Ham17}, $\nu$p-process \cite{Fro06, Pru06} the weak
r-process \cite{Qia07} (or sometimes it referred to as $\alpha$ process
\cite{Mey92}), or the $\gamma$-process \cite{Rau13}, may also contribute to
the observed abundance of the chemical elements and their isotopes. The last
two processes are particularly relevant for this study, since the predicted
abundances depend sensitively on the $\alpha$-nucleus optical potentials which
are used to derive the reaction rates. In the following we briefly introduce
these astrophysical scenarios. 

The most recent supernova simulations showed \cite{Hud10, Mar12} that the
neutrino-driven winds, emitted during the cooling of a neutron star born after
a massive star core collapse, are only slightly neutron-rich. As long as the
matter --- containing mainly protons and neutrons that form $\alpha$ particles
--- is relatively close to the neutron star, the high temperature maintains
the abundances in nuclear statistical equilibrium \cite{Woo92}. As the wind
expands, the temperature and the density decreases and the $\alpha$ particles
start to combine into heavier nuclei by ($\alpha$,n), (n,$\gamma$), (p,n), and
their inverse reactions. While these reactions are in statistical equilibrium,
the abundances along a given isotopic chain are determined by the neutron
density and the temperature of the environment. The lifetime of the produced
isotopes are long relative to the wind expansion and, therefore, $\beta$
decays are playing typically a marginal role only. However, as the temperature
drops below about T $\approx$ 4 GK the reactions, that are faster than the
$\beta$-decay of the produced isotopes --- such as ($\alpha$,n),
($\alpha,\gamma$), (p,$\gamma$) and (p,n) --- drive the matter towards heavier
masses. Under these conditions the nucleosynthesis path remains relatively
close to the valley of stability and light neutron-rich species between Fe and
Ag can be synthesized \cite{Arc11, Arc14, Moh16}.

On the other side of the valley of stability there are about 35 nuclei separated from the path of the neutron capture
processes. These mostly even-even isotopes between $^{74}$Se and $^{196}$Hg are the $p$ nuclei \cite{Rau13}. It is generally accepted that the main stellar mechanism synthesizing the $p$ nuclei -- the $\gamma$ process -- involves mainly photodisintegrations, dominantly ($\gamma$,n) reactions on preexisting more neutron-rich \textsl{s} and \textsl{r} seed nuclei. The high energy photons -- necessary for the $\gamma$-induced reactions -- are available in explosive nucleosynthetic scenarios where temperatures around a few GK are reached, like the Ne/O rich layer in core-collapse supernovae \cite{Woo78,Rau02} or during the thermonuclear explosion of a white dwarf (type Ia supernova) \cite{Tra11}. 
Recent work favors the latter scenario \cite{Nis18, Tra18}. Under these conditions, consecutive ($\gamma$,n) reactions drive the material towards the proton rich side of the chart of isotopes. As the neutron separation energy increases along this path, ($\gamma$,p) and ($\gamma,\alpha$) reactions become faster and process the material towards lighter elements \cite{Rau06, Rap06}. Theoretical investigations agree that ($\gamma$,p) reactions are more important for the lighter $p$ nuclei, whereas ($\gamma$,$\alpha$) reactions are mainly important at higher masses (neutron number $N\geq 82$) \cite{Rau13}.

The modeling of these two nucleosynthesis scenarios requires an extended
reaction network calculation involving several thousand reactions on mostly
unstable nuclei. However, the calculated abundances are very sensitive to the
applied reaction rates \cite{Rau06, Rap06, Bli17, Moh16} which are derived by
folding the reaction cross sections under stellar conditions with the
Maxwell-Boltzmann distribution at a given temperature. The cross sections are
predicted by the Hauser-Feshbach (H-F) statistical model \cite{hf} which
utilizes global optical model potentials (OMP). The nucleon-nucleus optical
potential is in general relatively well known, deviations between predictions calculated with different parameter sets are usually within factor of two, however, the rates calculated using different $\alpha$-nucleus optical potential parameter sets can disagree by even up to an order of magnitude \cite{Bli17, Kis13, Net13}. Typically, the
influence of the chosen level density parametrization remains minor, and the
$\gamma$-ray strength function mainly affects the ($\alpha$,$\gamma$) cross
section but has practically no impact on the ($\alpha$,n) cross
sections \cite{rauintjmod, rausensi}. Motivated by the $\gamma$ process nucleosynthesis, several cross
section measurements, mostly on proton-rich isotopes, were carried out in recent years to test the global $\alpha$-nucleus optical potentials \cite{Sau11, Sch14, Gyu14, Kis15} (and further references therein).

The aim of the present work is to evaluate the different open-access global and regional $\alpha$-nucleus potentials used in the weak r-process and $\gamma$-process network studies. As target nucleus we choose the $^{115}$In nucleus which lies only few mass units above the termination of the weak r process path. This isotope is relatively neutron-rich and furthermore, low energy $\alpha$-induced cross section data on odd-even nuclei is in general scarce. The cross sections of the $^{115}$In($\alpha,\gamma$)$^{119}$Sb, $^{115}$In($\alpha$,n)$^{118}$Sb$^m$ and $^{115}$In($\alpha$,n)$^{118}$Sb$^g$ reactions were measured and results are compared to theoretical predictions calculated by the TALYS code \cite{TALYS}. Moreover, the angular distributions of the $^{115}$In($\alpha,\alpha$)$^{115}$In elastic scattering were measured at energies around the Coulomb barrier recently \cite{Kis16}. 

This paper is organized as follows. Details on the experimental technique are
presented in Sec.~\ref{sec:exp}, and the experimental results summarized in
Sec.~\ref{sec:result}. The experimental data are compared to statistical
model calculations in Sec.~\ref{sec:theo} with a detailed presentation of the
global $\alpha$-nucleus optical potentials in Sec.~\ref{sec:glob} and a strict
$\chi^2$-based assessment in Sec.~\ref{sec:chi2}. The best-fit parameters of
$^{115}$In are used to predict $\alpha$-induced reaction cross sections for
the neighboring nucleus $^{113}$In in Sec.~\ref{sec:in113}. Conclusions are
drawn in Sec.~\ref{sec:sum}. Further details on the statistical model calculations are
provided in an Appendix.

\section{Experimental technique and results}
\label{sec:exp}

\begin{table*}
\caption{\label{tab:iso}Decay parameters of the Sb product nuclei taken from \cite{NNDC_Sb}. The yield of the gamma transition marked with $^+$ was sufficient for the analysis only at and above E$_{\alpha}$ = 14.0 MeV. Furthermore, the $\gamma$ transitions marked with $^*$ were used only at and above E$_{\alpha}$ = 15.0 MeV to determine the cross section.}
\begin{tabular}{ccccc}
\hline
\parbox[t]{1.2cm}{\centering{Nucleus}} &
\parbox[t]{1.2cm}{\centering{T$_{1/2}$ (hour)}} &
\parbox[t]{2.0cm}{\centering{Transition}} &
\parbox[t]{4.0cm}{\centering{x- and $\gamma$-ray \\energy (keV)}} &
\parbox[t]{4.0cm}{\centering{Relative intensity \\ per decay (\%)}} \\
\hline
$^{118}$Sb$^{g}$& 0.06 $\pm$ 0.0017& $\gamma$   &1229.3 & 2.5 $\pm$ 0.3 \\
           &                     & $\gamma$         &1267.2$^+$& 0.52 $\pm$ 0.07 \\
$^{118}$Sb$^m$& 5.00 $\pm$ 0.02  & K$_{\alpha2}$ &25.0 & 36.4 $\pm$ 1.2 \\
 					 &                     & K$_{\alpha1}$ &25.3 & 67.4 $\pm$ 2.2 \\
           &                     & $\gamma$         &40.8   & 30.2 $\pm$ 1.9 \\
					 &                     & $\gamma$         &253.7& 99 $\pm$ 6 \\
           &                     & $\gamma$         &1050.7$^*$& 99 $\pm$ 5 \\
           &                     & $\gamma$         &1229.7$^*$& 99 $\pm$ 5 \\
$^{119}$Sb &  38.19 $\pm$ 0.22   & K$_{\alpha2}$ &25.0 & 21.0 $\pm$ 0.5 \\
           &                     & K$_{\alpha1}$ &25.3 & 38.9 $\pm$ 0.9 \\
           &                     & $\gamma$         &23.9  & 16.5 $\pm$ 0.2 \\
\hline
\end{tabular}
\end{table*}

The element indium has two stable isotopes: $^{113}$In and $^{115}$In,
with natural abundances of 4.29 $\pm$ 0.04 \% and 95.71 $\pm$ 0.04 \%,
respectively \cite{NNDC_Sb}. Alpha capture and alpha-induced reactions on
$^{115}$In lead to unstable Sb isotopes, therefore, the cross sections
can be measured using the well-established activation method. However, the
half-lives of the reaction products range from 3.6 min to 1.6 days and
consequently two different $\gamma$ counting setups were used in the present
work to determine the reaction cross sections. Moreover, the ($\alpha$,n)
reaction on $^{115}$In populates the 
$1^+$
ground (T$_{1/2}$ = 3.6 min) and 
$8^-$
isomeric
(T$_{1/2}$ = 5.0 h) states in $^{118}$Sb. The isomeric state decays directly
to $^{118}$Sn without internal transition to the ground state. The decay
parameters of the reaction products are summarized in Table \ref{tab:iso}. In
the following chapters a detailed overview on target production and on the
counting setups will be given. 

\subsection{\label{sec:target} Target preparation and characterization}

Altogether twelve targets were prepared by vacuum evaporation of natural isotopic composition, high chemical purity (99.99\%) indium onto 2-2.5 $\mu$m thick Al backings and 0.5 mm thick Ta disks. During the evaporation, the backings were fixed in a holder placed 12 cm above the Ta evaporation boat. Due to the large distance between the evaporation boat and the backing it can be assumed that the surface of the targets is uniform. This assumption was checked by measuring the absolute target thickness at several spots using the Rutherford Backscattering Spectroscopy (RBS).  

\begin{center}
\begin{figure*}
\resizebox{1.80\columnwidth}{!}{\rotatebox{0}{\includegraphics[clip=]{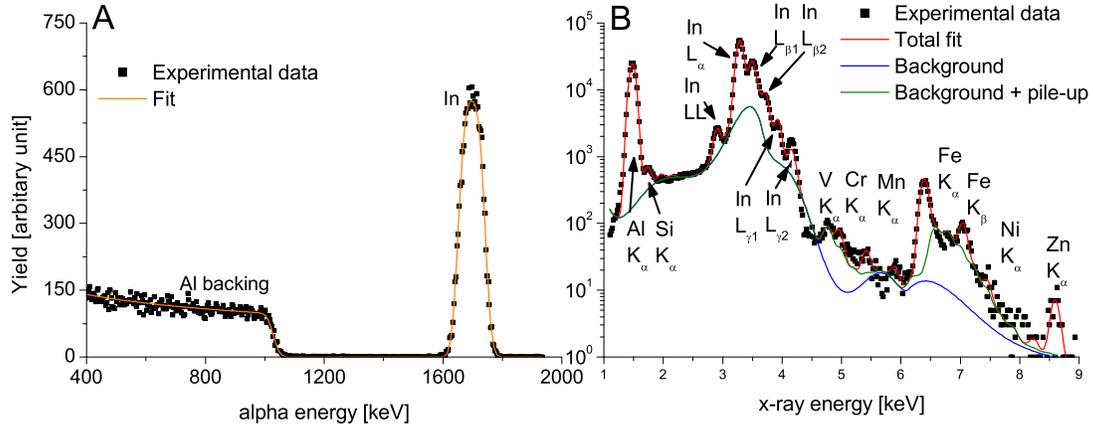}}}
\caption{\label{fig:target} Measured RBS (A) and PIXE (B) spectra used to characterize the targets. Peaks used for the analysis are
marked. Peaks belonging to impurities in the target and/or the backing are indicated too.}
\end{figure*}
\end{center}

The RBS measurements were performed using the Oxford-type Nuclear Microprobe Facility at Atomki, Debrecen, Hungary \cite{Hus16}. The energy of the $^4$He$^+$ beam provided by the Van de Graaff accelerator of Atomki was 1.6 MeV and 2 MeV. Two Silicon surface barrier detectors (50 mm$^2$ sensitive area and 18 keV energy resolution) were used to measure the yield of the backscattered ions, one of them was placed at a scattering angle of 165$^{\circ}$ (Cornell geometry) and the other one was set to 135$^{\circ}$ (IBM geometry). A typical RBS spectrum can be seen in Fig.~\ref{fig:target} A. The uncertainty of the target thickness determination carried out with the RBS technique is 3.0\% \cite{Hus16}.

The precise knowledge of the target (and backing) impurities is also of
crucial importance, because alpha-induced reactions on low mass impurities
could poison the measured gamma spectra. To characterize the targets, the
well-known particle induced x-ray emission (PIXE) technique was used
\cite{pixe}, too. The energy and the intensity of the proton beam provided by
the Van de Graaff accelerator of Atomki was 2 MeV and 1 nA, respectively. A
detailed description of the setup used can be found in \cite{Ker10}. A typical
PIXE spectrum can be seen in Fig.~\ref{fig:target} B. The following
impurities were found in the targets and in the backing used for the
alpha-induced cross section measurements (below 1 ppm): Si, V, Cr, Mn, Ni, Zn
and Ga, and (below 10 ppm) Fe. We assigned 4.0 \% uncertainty to the target thicknesses derived with the PIXE technique \cite{pixe}.

The absolute target thicknesses were found to be within 4.48 x 10$^{17}$ and 2.28 x 10$^{18}$ atoms/cm$^2$, the maximum deviation between the thicknesses determined by RBS and PIXE techniques were 4\%. Furthermore, the uniformity of the targets was
checked by measuring the backscattering alpha yield at several spots, the
maximum difference was found to be less than 2.7\%. Therefore, as a
conservative estimate, 4.8 \% has been adopted as the uncertainty of the
target thickness. 

\begin{center}
\begin{figure*}
\resizebox{1.80\columnwidth}{!}{\rotatebox{0}{\includegraphics[clip=]{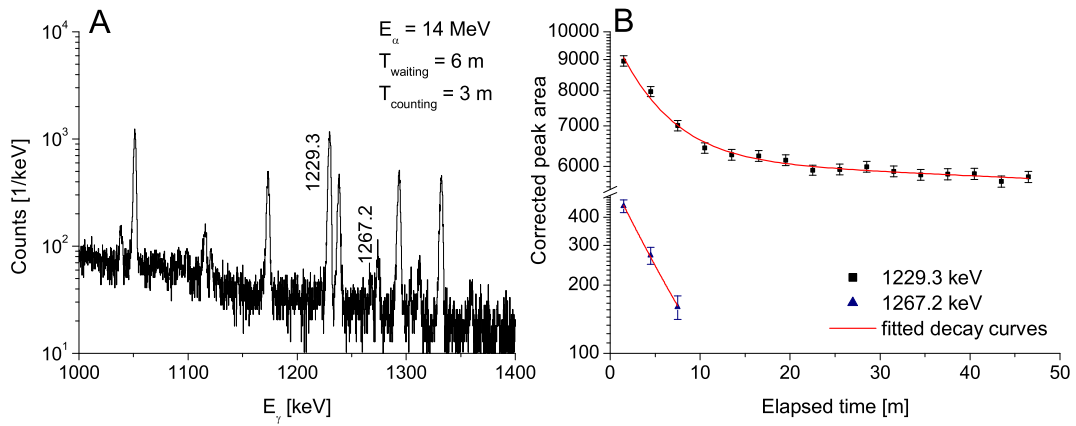}}}
\caption{\label{fig:gamma} Off-line $\gamma$-ray spectrum measured using Det1 (A), taken after irradiating an In target with E$_{\alpha}$ = 14.0 MeV beam. The peaks used for the
  analysis are marked. The other peaks belong to the decay of $^{116}$Sb (1293.6 keV) and
  $^{118}$Sb$^m$ (1050.7 keV) and to beam induced background --- $^{56}$Co
  (1238.3 keV), $^{60}$Co (1173.2 keV and 1332.5 keV). The deadtime and relative intensity corrected peak areas (decay curves) of
  the transitions used to determine the $^{115}$In($\alpha$,n)$^{118}$Sb$^{g}$
  reaction cross section is shown, too (B). The 1229.3 keV gamma-ray is emitted
  during the decay of both the produced $^{118}$Sb$^{g}$ and $^{118}$Sb$^{m}$
  isotopes. Therefore, the yield of this transition was fitted as the sum of
  two exponential functions in order to disentangle the two reaction
  channels. For further details see text.}
\end{figure*}
\end{center}

\begin{center}
\begin{figure*}
\resizebox{1.80\columnwidth}{!}{\rotatebox{0}{\includegraphics[clip=]{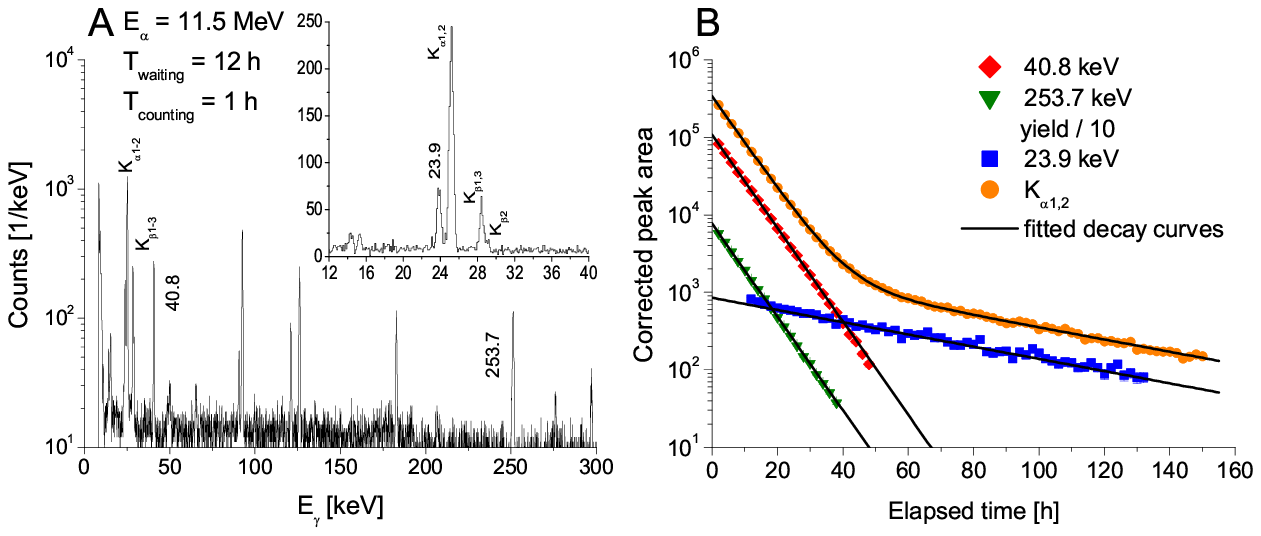}}}
\caption{\label{fig:gamma_leps} Off-line $\gamma$-ray spectra measured using the LEPS detector (A), taken after irradiating an In target with E$_{\alpha}$ = 11.5 MeV beam. The peaks used for the analysis are marked. The inset shows the spectrum taken 72 h after the end of the irradiation to highlight the transitions used to derive the alpha-capture cross sections. The deadtime and relative intensity corrected peak areas (decay curves) of the transitions used to determine the $^{115}$In($\alpha,\gamma$)$^{119}$Sb and $^{115}$In($\alpha$,n)$^{118}$Sb$^m$ reaction cross sections is shown, too (B). At the beginning of the counting the yield of the 23.9 keV transition was not sufficient for the analysis (due to the level of the background). For better visibility the yield of the 253.7 keV gamma line was scaled down by factor of 10 on the right side.}
\end{figure*}
\end{center}

\subsection{\label{sec:irrad} Irradiations}

The In targets were irradiated with $\alpha$ beams from the MGC20 cyclotron of
Atomki. The beam energies were in the range of 9.15 and 16.15 MeV, covered in
steps of about 0.5-1.15 MeV. After the beam-defining aperture, the chamber was
insulated and a secondary electron suppression voltage of -300 V was applied
at the entrance of the chamber. The typical beam current was between 0.6 $\mu$A and 0.8 $\mu$A, and
the length of each irradiation was between 0.25 h and 20 h, corresponding to
about 1.6 x 10$^{15}$ and 1.2 x 10$^{17}$ total incident $\alpha$
particles. Due to the short half-life of the produced $^{118}$Sb$^{g}$
isotope, shorter irradiations were used to determine the
$^{115}$In($\alpha$,n)$^{118}$Sb$^{g}$ reaction cross sections and longer ones
to measure the radiative alpha capture and the
$^{115}$In($\alpha$,n)$^{118}$Sb$^{m}$ reaction cross sections. 
During the irradiation, a Si ion-implanted detector --- built onto the wall of
the irradiation chamber at 165$^{\circ}$, with respect to the beam direction
--- was used to monitor the targets. Several beam tests were performed to
check the target stability before the experiment. These tests showed that
there is no deterioration, if the beam current is less than 1 $\mu$A. The
current integrator counts were recorded in multichannel scaling mode to take
into account the changes in the beam current. The scaler stepped channel in
every 10 sec. (in the case of the short irradiations) or 1 min. (in the case
of the irradiations aiming the measurement of the $^{115}$In($\alpha$,n)$^{118}$Sb$^{m}$ reaction cross
sections).  

\subsection{\label{sec:detectors} Determination of the absolute $\gamma$ detection efficiencies}

Three High Purity Germanium (HPGe) detectors were used to measure the yield of the emitted $\gamma$
rays: a 100\% relative efficiency coaxial HPGe (Det1), a 100\% relative efficiency
HPGe placed in low background shielding (Det2) and a Low Energy Photon
Spectrometer (LEPS) equipped with a 4$\pi$ multi-layer shielding including an inner 4 mm thick layer of copper, a 2 mm thick layer of cadmium, and an 8 cm thick outer lead shield \cite{Szu14}. To determine
the cross section of the radiative alpha capture and the
$^{115}$In($\alpha$,n)$^{118}$Sb$^{m}$ reactions the LEPS and Det2 were used. Det1 was used solely to study the
$^{115}$In($\alpha$,n)$^{118}$Sb$^{g}$ reaction. 

The distance between the source and the detector end cap during the $\gamma$
countings carried out with the LEPS detectors was 3 cm. To determine the
efficiency the following procedure was used: first the absolute detector
efficiency was measured at 15 cm distance from the surface of the LEPS detector
using $^{57}$Co, $^{133}$Ba, $^{152}$Eu, and $^{241}$Am calibrated sources. At
such large distance the so-called true coincidence summing effect is
negligible. Then, the activity of several $^{115}$In targets --- irradiated at
13 MeV, 14 MeV, 15 MeV and 16.15 MeV --- were measured in both close (3 cm)
and far (15 cm) geometries. Taking into account the time elapsed between the
two countings, a conversion factor of the efficiencies between the two
geometries could be determined and used henceforward in the analysis. To limit
the possible systematic uncertainty of our data, the activity of a few
irradiated samples were measured using Det2, too. These $\gamma$ countings were performed solely in far
geometry: the targets were placed 27 cm away from the end cap of the detector.

The absolute detector efficiency of Det1 and Det2 was derived in far geometry by using $^{60}$Co
and the aforementioned sources. To determine the far-to-close geometry
conversation factor for Det1, three $^{115}$In targets
were irradiated at 16.15 MeV and the emitted $\gamma$ yields were measured
with the detector in far geometry. After 3 hours of waiting time the
irradiations and $\gamma$ countings were repeated for each target, however,
this time the detector was positioned into close geometry. The waiting times
and the lengths of the irradiations were the same. The difference in the
number of impinging $\alpha$ particles were below 1.4\% and this factor was
taken into account by using the multichannel scaling spectrum. After
subtracting the contribution to the yield of the 1229.3 keV $\gamma$ line
originated from the decay of the $^{118}$Sb$^m$, the ratio of the number of
$\gamma$'s measured in the same time interval at far and close distances
correspond to the ratio of the far and close efficiencies. This approach
was repeated for all three samples counted at both far and close geometry
and an efficiency conversation factor was determined.

\subsection{\label{sec:short} Gamma counting with Det1}

The half-life of the produced $^{118}$Sb$^{g}$ is only 3.6 min, therefore, our
usual approach - described e.g. in \cite{Yal09} - cannot be used. Instead, we
placed Det1 in the experimental hall near the irradiation chamber and kept there for the cross section
measurement. After the irradiations, 0.5 min waiting time was used in order to
let short-lived activities decay which would impact the quality of the
measurement. The duration of the countings were about 45 min in case of each
irradiation.

The produced $^{118}$Sb$^{g}$ decays via the emission of several weak
$\gamma$-rays with relative intensities below 2.5\%. To determine the reaction
cross section the yield of the E$_{\gamma}$ = 1229.3 keV $\gamma$-ray was
measured using Det1. However, the
1229.3 keV $\gamma$ ray does not belong solely to the decay of the
$^{118}$Sb$^{g}$, a similar energy (E$_{\gamma}$ = 1229.7 keV) $\gamma$-ray is
emitted also after the electron capture decay of the $^{118}$Sb$^{m}$. Therefore,
to disentangle the two different reaction channels, each $\gamma$ counting
lasted for at least 45 min (the spectra was stored in every min.) and the
measured yields as the function of elapsed time was fitted with the sum of two
exponential functions with known half-lives. Above E$_{\alpha}$ = 14 MeV the
yield of the emitted 1267.2 keV $\gamma$-ray was sufficient for the
cross section determination, too. Figure \ref{fig:gamma} shows a typical
$\gamma$ spectrum (A) measured with Det1 and on the right side
the decay curve (B) of the transitions used to determine the
$^{115}$In($\alpha$,n)$^{118}$Sb$^{g}$ reaction cross sections. It has to be
mentioned that the cross sections based on the counting of the 1229.3 keV and
1267.2 keV $\gamma$-rays were always within their statistical uncertainties. 

\subsection{\label{sec:long} Gamma counting performed using the Low Energy Photon Spectrometer}

\begin{center}
\begin{figure}
\resizebox{1.0\columnwidth}{!}{\rotatebox{0}{\includegraphics[clip=]{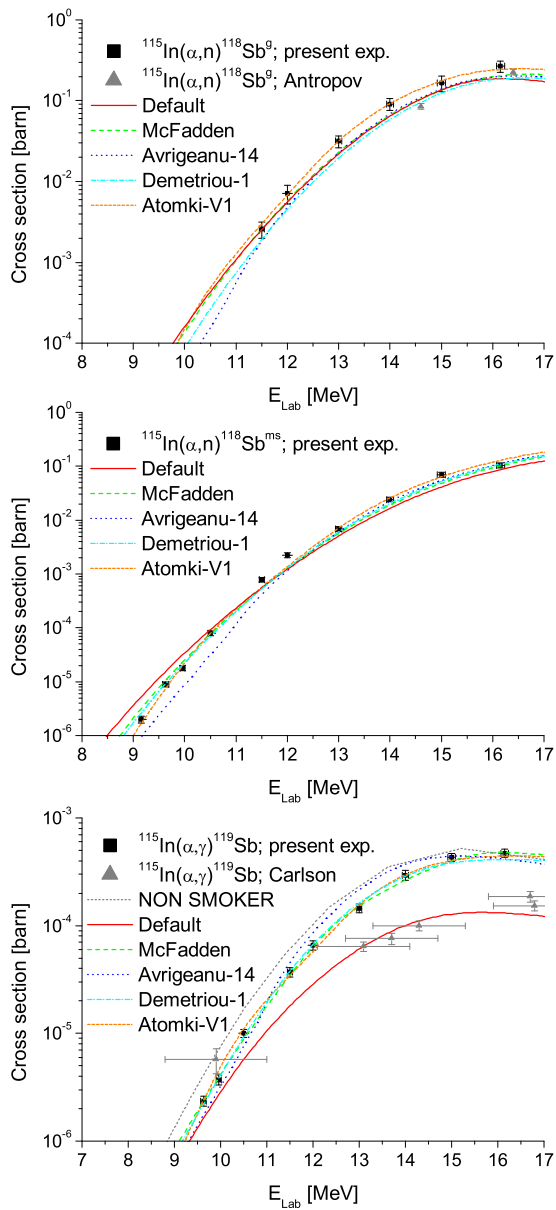}}}
\caption{\label{fig:exp_st_th} Experimental cross sections of the $^{115}$In($\alpha$,n)$^{118}$Sb$^g$,
  $^{115}$In($\alpha$,n)$^{118}$Sb$^m$ and $^{115}$In($\alpha,\gamma$)$^{119}$Sb reactions compared to calculations 
  using different combinations of input parameters of the statistical
  model. The curves --- calculated using the TALYS code \cite{TALYS} --- are labeled by the most important parameter which is the
  $\alpha$-nucleus OMP. For completeness the ($\alpha,\gamma$) cross section predictions, calculated with the widely used NON SMOKER code \cite{non} is plotted, too. For more details see text.}
\end{figure}
\end{center}

\begin{center}
\begin{figure*}
\resizebox{1.80\columnwidth}{!}{\rotatebox{0}{\includegraphics[clip=]{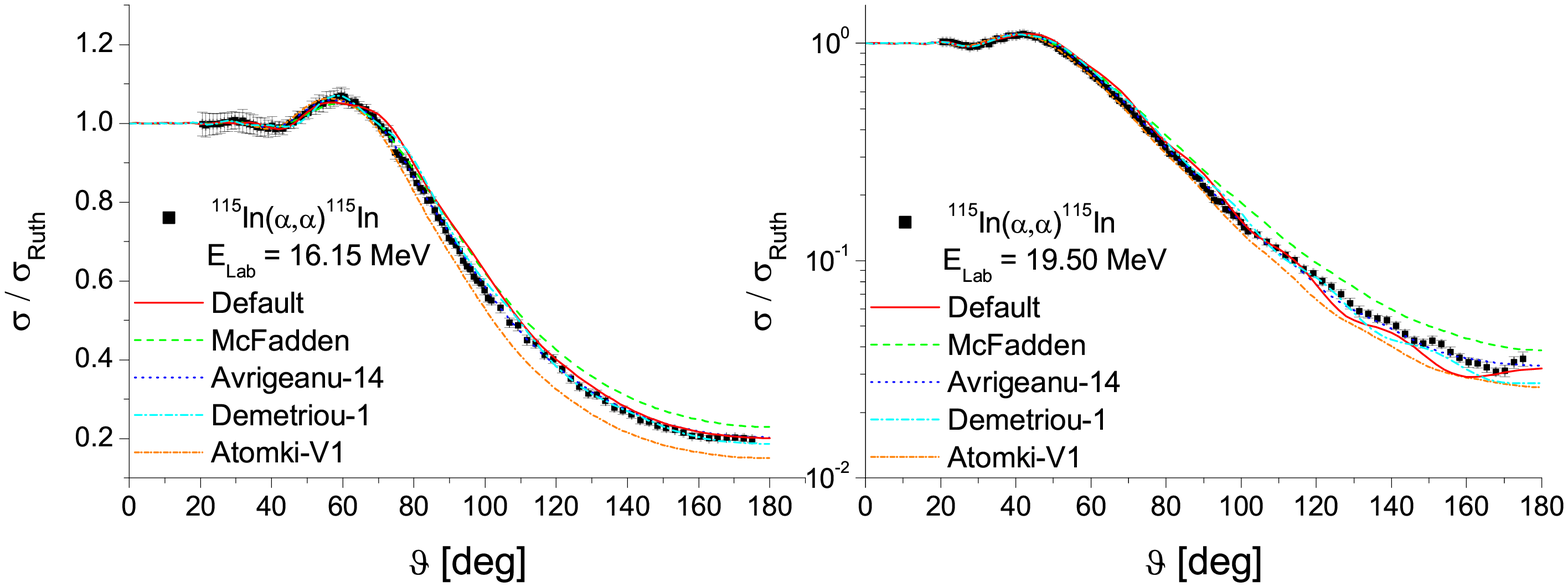}}}
\caption{\label{fig:an_di} Rutherford normalized elastic scattering cross sections of the $^{115}$In($\alpha,\alpha$)$^{115}$In reaction at E$_{Lab}$ = 16.15 and 19.50 MeV versus the angle in center-of-mass frame. The lines correspond to the predictions calculated using the global optical potential parameter sets discussed in chapter III/A.}
\end{figure*}
\end{center}

The characteristic x-ray and $\gamma$ radiation emitted during the electron-capture decay of the produced $^{119}$Sb and $^{118}$Sb$^m$ isotopes was measured using a LEPS. 

The countings were started typically 2 h after the end of the irradiation and lasted for 100 to 200 h. During the counting the spectra were stored
regularly (every hour) to follow the decay of the reaction
products. Figure~\ref{fig:gamma_leps} shows typical spectra measured with the
LEPS detector (A) and the decay curve of the $\gamma$-rays emitted during the
$\beta$-decay of the produced Sb reaction products (B). In the case of all
irradiated samples the source activities have been determined using the
$\gamma$- and characteristic x-ray transitions listed in Table \ref{tab:iso},
consistent results were obtained and their statistical uncertainty weighted averages were adopted as
the final results. 

\begin{table*}
\caption{\label{tab:xs} Measured cross sections of the $^{115}$In($\alpha,\gamma$)$^{119}$Sb, $^{115}$In($\alpha$,n)$^{118}$Sb$^m$ and $^{115}$In($\alpha$,n)$^{118}$Sb$^g$ reactions.}
\begin{tabular}{ccccc}
\hline
\hline
\parbox[t]{1.4cm}{\centering{E$_{lab}$ \\  (MeV)}} &
\parbox[t]{1.4cm}{\centering{E$_{c.m.}$ \\ (MeV)}} &
\parbox[t]{3.4cm}{\centering{$^{115}$In($\alpha,\gamma$)$^{119}$Sb \\ cross section  ($\mu$barn)}} &
\parbox[t]{3.4cm}{\centering{$^{115}$In($\alpha$,n)$^{118}$Sb$^m$ \\ cross section (mbarn)}} &
\parbox[t]{3.4cm}{\centering{$^{115}$In($\alpha$,n)$^{118}$Sb$^{g}$ \\ cross section (mbarn)}} \\
\hline
9.15	&8.83	  $\pm$ 0.04	&0.89	$\pm$	0.11 & 0.002 $\pm$ 0.0003 & \\
9.63	&9.28	  $\pm$ 0.06	&2.36	$\pm$	0.25 & 0.009 $\pm$ 0.001  & \\
9.96	&9.61	  $\pm$ 0.05	&3.7	$\pm$	0.4	 & 0.018 $\pm$ 0.0018 & \\
10.50	&10.14	$\pm$ 0.04	&10.0 $\pm$	0.9	 & 0.080 $\pm$ 0.008  & \\
11.50	&11.10	$\pm$ 0.05	&37		$\pm$ 3.5	 & 0.78	 $\pm$ 0.08   &   2.57 $\pm$ 0.57\\
12.00	&11.59	$\pm$ 0.04	&66		$\pm$ 6.7	 & 2.22  $\pm$ 0.23   &   7.16 $\pm$ 1.85\\
13.00	&12.54	$\pm$ 0.06	&145	$\pm$	13	 & 6.85	 $\pm$ 0.66   &   31.6 $\pm$ 5.3\\
14.00 &13.49  $\pm$ 0.07  &295  $\pm$ 31   & 24.1  $\pm$ 2.3    &   91.4 $\pm$ 15.0\\ 
15.00 &14.48  $\pm$ 0.08  &431  $\pm$ 39   & 70.2  $\pm$ 7.1    &    165 $\pm$ 36\\ 
16.15 &15.58  $\pm$ 0.08  &470  $\pm$ 41	 & 104   $\pm$ 10.8   &    267 $\pm$ 44\\			
\hline
\end{tabular}
\end{table*}

To determine the cross section of the $^{115}$In($\alpha,\gamma$)$^{119}$Sb reaction, at first the 23.9 keV $\gamma$ transition was used. In the case of the $^{115}$In($\alpha$,n)$^{118}$Sb$^m$ reaction, the yield of the 40.8 keV and 253.678 keV transitions were measured using the LEPS. 

The activity of the samples irradiated at E$_{c.m.}$ = 14.48 and 15.58 MeV was derived also using Det2. To determine the cross section of the $^{115}$In($\alpha$,n)$^{118}$Sb$^m$ the yield of the 1050.7 and 1229.7 keV $\gamma$ rays were measured in far geometry. The resulted cross sections were within the statistical uncertainty compared to the ones derived using the LEPS. Furthermore, the yield of the characteristic x-ray transitions were used also to derive the cross sections using the procedure described below.

The energies of the emitted characteristic x-ray K$_{\alpha1,2}$ lines are 25.0 and 25.3 keV, respectively. Because of the
resolution of the LEPS detector is typically between 400 eV (for a 5.9 keV line) and 680 eV (for a 122 keV ray), in the
x-ray spectra it is not possible to distinguish between K$_{\alpha1}$ and
K$_{\alpha2}$ transitions. Instead, the sum of the emitted characteristic x-rays was used for the analysis. 
At the beginning of the counting the characteristic x-ray yield is dominated by the decay of the $^{118}$Sb$^m$ nuclei, however, after about 2 days (depending on the beam energy) in the x-ray peaks the decay of the $^{119}$Sb becomes dominant. We fitted the characteristic x-ray decay curve with a sum of two exponentials with the known half-lives to derive the number of reaction products. 
The agreement between the alpha-induced cross sections based on the $\gamma$- and x-ray counting were within 4\%, better than the independent uncertainties of the two values.

\subsection{\label{sec:result} Cross section results}
The final results of the present study are listed in Table \ref{tab:xs}, and a
comparison to statistical model calculations using the codes TALYS
\cite{TALYS} and NON-SMOKER \cite{non} is shown in
Fig.~\ref{fig:exp_st_th}. The quoted uncertainty in the E$_{c.m.}$ values
corresponds to the energy calibration of the alpha beam and to the uncertainty
of the energy loss in the target, which was calculated using the SRIM code
\cite{SRIM}. Several irradiations were repeated, in these cases the cross sections
were derived from the averaged results of the irradiations weighted by
the statistical uncertainty of the measured values. The uncertainty of the
cross sections is the quadratic sum of the following partial errors:
efficiency of the detectors determined at far geometry (5\% for all
detectors), close geometry efficiency correction factor (0.8\% for the LEPS
and 2.2\% for Det1 and Det2), number of target atoms (4.8\%), current
measurement (3\%), uncertainty of decay parameters ($\leq$ 6\%), and counting
statistics (0.5\% - 16.8\%).

\subsection{Experimental data from literature}
\label{sec:other}
The cross section of the $^{115}$In($\alpha,\gamma$)$^{119}$Sb reaction was measured by Carlson \textit{et al.,} in the middle of the 1960`s \cite{Car67} using radio-chemical techniques and NaI(Tl) detectors. The few mg/cm$^2$ $^{115}$In targets were made using the Zapon painting technique \cite{Zap52} and the thicknesses were determined with an estimated $\pm$ 10\% accuracy by weight and area measurements. The errors of the E$_{c.m.}$ --- shown in Fig. \ref{fig:exp_st_th} --- are the sum of the $\pm$ 500 keV initial energy spread of the cyclotron beam and the energy straggling in the foil stacks. The cross section of the $^{115}$In($\alpha$,n)$^{118}$Sb and $^{115}$In($\alpha$,2n)$^{117}$Sb was measured by A. E. Antonov \cite{Ant91} between 14.6 MeV $\leq$ E$_{lab.}$ $\leq$ 24.1 MeV (slightly above the region covered by our experimental data) and 16.4 MeV $\leq$ E$_{lab.}$ $\leq$ 23.0 MeV, respectively. Unfortunately no refereed publication on the measurement is available, the data were published solely in a conference proceedings. Furthermore the cross sections were measured only at very few (7 and 5) energies and information on the E$_{lab.}$ uncertainties is not available in \cite{Ant91}. For completeness both the ($\alpha,\gamma$) and ($\alpha$,n) data of \cite{Car67} and \cite{Ant91} are shown in Fig \ref{fig:exp_st_th}, however, their precision is not at the required level and, therefore, we omit these data from the detailed analysis presented in the following chapters.

\section{Statistical model calculations}
\label{sec:theo}
In general, the $\alpha$-nucleus OMP is identified as the most important
ingredient for the statistical model calculations (see Sec.~\ref{sec:chi2}
below) because it defines the total $\alpha$-induced reaction cross section
$\sigma_{\rm{reac}}$. Therefore we start with a presentation of $\alpha$-OMPs
which are widely used for the calculation of low-energy reaction cross
sections. These potentials are included in the statistical model code TALYS
(V.~1.8) except the Atomki-V1 potential \cite{Moh_ADNDT}. As the source code
of TALYS is available, the Atomki-V1 potential was implemented as an
additional option. However, in Fig \ref{fig:exp_st_th} for completeness the
($\alpha,\gamma$) cross section predictions, calculated with the widely used
NON SMOKER code \cite{non} is plotted, too. While this code provides good
description at high energies, in Fig \ref{fig:exp_st_th} it is clearly shown
that toward lower energies the discrepancy between the experimental data and
the predictions are increasing reaching roughly a factor of two at the lowest
measured cross sections.

\subsection{\label{sec:glob} Global $\alpha$-nucleus optical potentials and the new experimental data}

In the framework of the weak r process and $\gamma$ process network
calculations a large number of reactions involving --- either in the entrance
or in the exit channel --- $\alpha$-particles has to be taken into account. As
the path of these processes is located in a region of unstable nuclei
experimental data are practically not available to adjust the parameters of
the $\alpha$+nucleus potential. Therefore, a global $\alpha$+nucleus optical
potential is required for the theoretical prediction of reaction cross
sections. Several different parameter sets for the optical potential exist,
giving very different predictions for reaction cross sections in particular at
very low energies far below the Coulomb barrier \cite{Kis13, Bli17}.
In the following we will compare the predictions calculated using well known
or recent open access global potentials to the experimental results. The
reaction and scattering cross section calculations were performed using the
TALYS code (version 1.8) \cite{TALYS} with an extension for the Atomki-V1
potential.

($i$) The optical model potential calculations within TALYS are performed
with ECIS-2006 \cite{ECIS} using a default OMP based on a
simplification of the folding approach of Watanabe \cite{Wat58}. (Note that
this default choice for the $\alpha$-OMP in TALYS will change to
\cite{Avr10,Avr14} in later versions.)

($ii$) Elastic $\alpha$ scattering experiments at E$_{\alpha}$=24.7 MeV on
nuclei ranging from O to U were performed in the middle of the 1960's and a
consistent optical potential study was carried out. This study resulted in the
widely used global $\alpha$-nucleus optical model potential of McFadden and
Satchler \cite{Mcf}. The potential itself is very simple, it consists of a
4-parameter Woods-Saxon potential with mass- and energy-independent
parameters. Due to its simplicity this potential is used as default for the
H-F calculations of astrophysical reaction rates in the widely used NON-SMOKER
code \cite{Rau00,Rau01}.
 
($iii$) The real part of the potential of Demetriou, Grama and Goriely
\cite{Dem02} is described in the framework of the double-folding model of
Kobos \textsl{et al}. \cite{Kob84}. While the shape of the potential is
directly described by the double-folding procedure, its strength is adjusted
according to the available experimental data. A simple Woods-Saxon form is
used to describe the imaginary part of the optical model potential, however,
three different parameter sets are introduced. Namely, potential I consists of
a volume term only, potential II combines a volume and surface component and
finally, potential III, the “dispersive optical model” relates the real and
imaginary part of the optical model potential through the dispersion
relation. We found that --- among the \cite{Dem02} potentials --- model I
provides the best description for the cross sections and angular
distributions, therefore we show in Fig. \ref{fig:exp_st_th} and
Fig. \ref{fig:an_di} only the results calculated using this OMP.

($iv$) The $\alpha$-nucleus optical potential of M. Avrigeanu \textsl {et
  al}. \cite{Avr10} was obtained by fitting elastic $\alpha$ particle
scattering angular distributions and reaction cross sections on nuclei with
masses between 45 $\leq$ A $\leq$ 209 and for E $\leq$ 50 MeV. The potential
consists of Woods-Saxon parametrizations for both the real and the imaginary
parts. The real part is characterized by three parameters (depth V$_0$, radius
R$_{R}$ and diffuseness a$_R$), the imaginary part is described using the sum
of volume and a surface potential, determined by 6 parameters (W$_V$, R$_V$,
A$_{v}$; W$_S$, R$_S$, a$_S$) \cite{Avr10}. All of nine parameters are mass
and energy dependent this way this potential provides an excellent description
for both the reaction and scattering data. However, it has to be emphasized
that the extrapolation of a many-parameter potential into the region of
unstable nuclei may lead to additional uncertainties in the calculation of
astrophysical reaction rates. An updated version of this potential was
published recently \cite{Avr14} and this modified potential is used in the
present study.

($v$) Several $\alpha$ elastic scattering experiments on the target nuclei
$^{89}$Y, $^{92}$Mo, $^{106,110,116}$Cd, $^{113,115}$In, $^{112,124}$Sn, and
$^{144}$Sm have been performed at Atomki in recent years \cite{Kis09, Ful01,
  Kis06, Orn15, Kis11, Kis13, Kis16, Gal05, Moh97}.  Based solely on these high
precision elastic $\alpha$ scattering data, a new few-parameter global optical
potential parametrization --- so-called Atomki-V1 --- has been suggested in
\cite{Moh_ADNDT}. The geometry of the energy-independent real part of the
potential is determined using the folding procedure. The imaginary part of the
potential is described by surface Woods-Saxon potential with
energy-independent radius and diffuseness parameters. This potential gives a
correct prediction for the total $\alpha$-induced reaction cross sections
\cite{Moh_tot} and, furthermore, the very few adjustable parameters avoid
contingent problems which may appear in the extrapolation of many-parameter
potentials for unstable nuclei.

\subsection{$\chi^2$-based assessment of the calculations}
\label{sec:chi2}
The cross section of an ($\alpha$,$X$) reaction in the statistical model
depends on the total transmission coefficients $T_i$ into the open channels.
(Note that the total transmission and average width for a particular channel
are closely related, see e.g. Eq.~(64) and Eq.~(65) in
\cite{rauintjmod,rausensi}):
\begin{equation}
\sigma(\alpha,X) \propto \frac{T_{\alpha,0} \, T_X}{\sum_i T_i} =
T_{\alpha,0} \times b_{X}
\label{eq:HF}
\end{equation}
with the branching ratio into the $X$ exit channel $b_X = T_X/\sum_i T_i$.
The strict application of Eq.~(\ref{eq:HF}) does not take into account a
preequilibrium contribution. This choice will be discussed later.

In the case of $\alpha$-induced reactions on $^{115}$In at energies
sufficiently above the neutron-threshold, we find that
$\sigma(\alpha,\mathrm{n}) \propto T_{\alpha,0}$ and $\sigma(\alpha,\gamma)
\propto T_{\alpha,0} T_\gamma / T_\mathrm{n}$ \cite{rausensi}. Consequently,
$\sigma(\alpha,\mathrm{n})$ is essentially defined by the $\alpha$ potential,
and experimental data can be used to constrain the $\alpha$ potential. As soon
as $T_{\alpha,0}$ is fixed, $\sigma(\alpha,\gamma)$ provides a constraint for
the ratio $T_\gamma/T_\mathrm{n}$ but it is not possible to determine
$T_\gamma$ or $T_\mathrm{n}$ individually. The experimental data close above the
($\alpha$,n) threshold provide some information on $T_p$.

Note that the $T_i$ result from the sum over all energetically accessible
final states in the respective channel. In practice, this sum is typically
composed of the several known low-lying states which are considered explicitly
plus the contributions of high-lying states which are calculated from a level
density parametrization.

Consequently, the calculated reaction cross sections depend on four physical
input parameters which are the $\alpha$-nucleus potential (see above), the
nucleon-nucleus potential, the $\gamma$-ray strength function (composed of E1
and M1 components), and the level density. TALYS \cite{TALYS} provides 8
different options for the $\alpha$-potential which is the basic ingredient for
the calculation of ($\alpha$,$X$) cross sections. In addition, we have
implemented the Atomki-V1 potential in TALYS, and we have used scaling factors
between 0.7 and 1.2 for the Demetriou potential as suggested by \cite{Sch16}.

For a $\chi^2$-based assessment of the TALYS inputs, excitation functions for
the $^{115}$In($\alpha,\gamma$)$^{119}$Sb,
$^{115}$In($\alpha$,n)$^{118}$Sb$^{g}$, and
$^{115}$In($\alpha$,n)$^{118}$Sb$^m$ reactions were calculated from all
combinations of the TALYS input parameters. In total, from the 6720
combinations of input parameters $\chi^2$ values (all $\chi^2$ are given per
point in the following) between 1.7 and more than 30000 were found,
corresponding to average deviation factors (f$_{dev}$ in the following)
between 1.12 in the best cases and up to a factor of about 6 in the worst
cases. Details of this procedure are given in \cite{Mohr17}.

In general, the best description of all reaction cross sections is obtained
with $\chi^2 < 2$ from the Atomki-V1 potentials. Slightly increased $\chi^2
\approx 3$ are achieved from the first version of the Demetriou potentials
(Dem-1) and from the simple McFadden/Satchler potential. The other
$\alpha$-potentials show best fits of $\chi^2 \approx 10$ or
higher. Surprisingly, this holds also for the latest potential of Avrigeanu
{\it et al.}\ \cite{Avr14} which provides an excellent description of the
$^{115}$In($\alpha$,$\alpha$)$^{115}$In elastic scattering data (see
Sec.~\ref{sec:scat}). Besides the total $\chi^2$ for all reaction channels, we
now investigate also the different exit channels separately.

The $^{115}$In($\alpha$,n)$^{118}$Sb$^{g}$ data can be described very well in
their restricted energy range with $\chi^2 \approx 0.1$ ($f_{\rm{dev}} \approx
1.06$) using the Atomki-V1 potential. Several other potentials reach small
$\chi^2 \lesssim 2$ ($f_{\rm{dev}} \lesssim 1.3$): these values are achieved
from several versions of the Demetriou potentials, the McFadden/Satchler
potential, the Avrigeanu potential, and the Watanabe potential. Contrary to
the excellent reproduction of the ground state contribution, the
$^{115}$In($\alpha$,n)$^{118}$Sb$^m$ isomer production cannot be described
over the whole energy range with a similar precision: the best description
with $\chi^2 \approx 2.7$ and $f_{\rm{dev}} \approx 1.18$ is
found for the Atomki-V1 potential; slightly worse numbers ($\chi^2 \approx 4 -
4.5$ and $f_{\rm{dev}} \approx 1.2$) are obtained for the Dem-1 and the
McFadden/Satchler potentials. 

Because of the sensitivity of the $^{115}$In($\alpha,\gamma$)$^{119}$Sb cross
section to a combination of input parameters, it is not possible to constrain
any particular parameter from the $(\alpha$,$\gamma$) data alone. Several
combinations of parameters (including combinations with overall very poor
$\chi^2$ values reaching values above 100!) are able to reproduce the
($\alpha$,$\gamma$) data alone with $\chi^2 \lesssim 1.5$. The best
combinations of input parameters are able to reach $\chi^2 \approx 0.5$ and
$f_{\rm{dev}} \approx 1.05$; the corresponding $\alpha$-nucleus potentials are
the McFadden/Satchler and the Dem-2 potentials.

Obviously, it is impossible to include all 6720 excitation functions for
presentation in Fig.~\ref{fig:exp_st_th}. The following selection of
excitation functions is labeled in Fig.~\ref{fig:exp_st_th} by the most
important ingredient which is the $\alpha$-nucleus OMP. These examples
are also summarized in Table \ref{tab:chi2}.

\begin{table*}
\caption{
\label{tab:chi2} 
$\chi^2$ and $f_{\rm{dev}}$ for some selected combinations of
input parameters for the statistical model. For further discussion see text.
} 
\setlength{\extrarowheight}{0.1cm}
\begin{ruledtabular}
\begin{tabular}{ccccrrrrrrrr}
\multicolumn{4}{c}{parameters}
& \multicolumn{2}{c}{all reactions}
& \multicolumn{2}{c}{($\alpha$,$\gamma$)}
& \multicolumn{2}{c}{($\alpha$,n)-gs}
& \multicolumn{2}{c}{($\alpha$,n)-iso} \\
$\alpha$-OMP 
& nucleon-OMP
& $\gamma$-ray strength
& level density
&
$\chi^2$ & $f_{\rm{dev}}$ &
$\chi^2$ & $f_{\rm{dev}}$ &
$\chi^2$ & $f_{\rm{dev}}$ &
$\chi^2$ & $f_{\rm{dev}}$ \\
\hline
WAT \footnote{TALYS default}      
& KD \footnotemark[1]
& KU \footnotemark[1]
& CTM \footnotemark[1]
& 307.8    & 1.89    
& 785.5    & 3.03    
&   2.1    & 1.33    
&  13.7    & 1.54    
\\
Atomki-V1 \footnote{overall best-fit}      
& JLM0-B \footnotemark[2]
& HFB \footnotemark[2]
& BSFG \footnotemark[2]
&   1.7    & 1.12    
&   0.8    & 1.08    
&   0.1    & 1.07    
&   3.4    & 1.19    
\\
Dem-1 
& JLM2-BG
& HG 
& MGH
&   2.8    & 1.23    
&   0.8    & 1.07    
&   3.2    & 1.50    
&   4.7    & 1.24    
\\
McF
& JLM1-BG
& TRMF 
& MSG
&   3.3    & 1.21    
&   0.8    & 1.08    
&   1.7    & 1.27    
&   6.8    & 1.31    
\\
AVR
& JLM3-BG
& HG 
& MSHC
&   9.7    & 1.45    
&   6.8    & 1.25    
&   2.1    & 1.38    
&  17.2    & 1.75    
\\
\hline
\multicolumn{12}{l}{\parbox[t]{0.8\textwidth}{\raggedright
$\alpha$-OMPs: 
Watanabe: WAT \cite{Wat58}; 
Atomki-V1 \cite{Moh_ADNDT};
Demetriou: Dem-1 \cite{Dem02}; \\
\hspace{0.5cm}
McFadden/Satchler: McF \cite{Mcf};
Avrigeanu: AVR \cite{Avr14}.
}} \\
\multicolumn{12}{l}{\parbox[t]{0.8\textwidth}{\raggedright
nucleon-OMPs: 
Koning-Delaroche: KD \cite{Kon03};
Jeukenne-Lejeune-Mahaux-Bauge: JLM0-B \cite{Bau01}; \\ 
\hspace{0.5cm}
Jeukenne-Lejeune-Mahaux-Bauge-Goriely: JLM2-BG \cite{Bau01,Gor07}; \\
\hspace{0.5cm}
Jeukenne-Lejeune-Mahaux-Bauge-Goriely: JLM1-BG \cite{Bau01,Gor07}; \\
\hspace{0.5cm}
Jeukenne-Lejeune-Mahaux-Bauge-Goriely: JLM3-BG \cite{Bau01,Gor07}.
}} \\
\multicolumn{12}{l}{\parbox[t]{0.8\textwidth}{\raggedright
$\gamma$ strengths: 
Kopecky-Uhl: KU \cite{Kop90};
Hartree-Fock-Bogolyubov: HFB \cite{Gor04};
Hybrid-Goriely: HG \cite{Gor98}; \\
\hspace{0.5cm}
Temperature-dependent Relativistic Mean Field: TRMF \cite{Cap09}.
}} \\
\multicolumn{12}{l}{\parbox[t]{0.8\textwidth}{\raggedright
level densities: 
Constant Temperature Model: CTM \cite{Gil65}; 
Back-Shifted Fermi Gas: BSFG \cite{Dilg73}; \\ 
\hspace{0.5cm} 
Microscopic Gogny Hilaire: MGH \cite{Hil12};
Microscopic Skyrme Goriely: MSG \cite{Gor01}; \\
\hspace{0.5cm}
Microscopic Skyrme Hilaire Combinatorial: MSHC \cite{Gor08}.

}} \\
\end{tabular} 
\end{ruledtabular}
\end{table*}

($i$) The result from the TALYS default parameters is based on the early
$\alpha$-nucleus OMP by Watanabe \cite{Wat58}. The default calculation
provides reasonable description of the ($\alpha$,n) data, 
but the energy dependence of the isomer cross section is not well reproduced.
This holds also for the ($\alpha$,$\gamma$) data which are significantly
underestimated at higher energies.
This clearly points to a deficiency of the default $\alpha$-OMP, the default
$\gamma$-ray strength function and/or the default level
density. Interestingly, by accident the default calculation matches the
earlier $(\alpha$,$\gamma$) data of Carlson {\it et al.}\ \cite{Car67} which
are much lower than the present results.

($ii$) The overall best-fit calculation is obtained from a combination of the
Atomki-V1 $\alpha$-OMP \cite{Moh_ADNDT}, a JLM-type (Jeukenne, Lejeune, and
Mahaux, e.g.\ \cite{JLM}) nucleon OMP in the version of Bauge {\it et
  al.}\ \cite{Bau01} (so-called ``{\it{jlmmode 0}}''), the microscopic
Hartree-Fock Bogolyubov E1 $\gamma$-ray strength function of Goriely
\cite{Gor04}, the M1 $\gamma$-ray strength normalized to the E1 strength
(TALYS default), and the back-shifted Fermi gas level density 
\cite{Dilg73}. 

($iii$) Among the different versions of the Demetriou potentials \cite{Dem02},
including scaling factors between 0.7 and 1.2 for the third version, the first
version (Dem-1) shows the best agreement with the experimental data in
combination with again a JLM type nucleon OMP, now of so-called ``{\it{jlmmode
    2}}'' \cite{Bau01,Gor07}, and the microscopic level density by Hilaire
\cite{Hil12}. In general, the ($\alpha$,n) cross sections are slightly
underestimated.
In the $(\alpha$,$\gamma$) channel
these relatively low cross sections from the Dem-1 potential can be
compensated by a larger E1 strength using Goriely's hybrid model \cite{Gor98}.

($iv$) The best result for the McFadden/Satchler potential describes the
experimental data with the same quality as the above Dem-1 potential. It is
obtained in combination with a JLM-type nucleon OMP (``{\it{jlmmode 1}}'')
\cite{Bau01,Gor07}, the microscopic level density from Skyrme forces
\cite{Gor01}, and the temperature-dependent relativistic mean-field E1
strength \cite{Cap09}.
It may be noted that a special combination of the McFadden/Satchler potential
with the other parameters close to the above Atomki-V1 best-fit provides the
best description of the ($\alpha$,$\gamma$) channel with $\chi^2 \approx 0.4$
and an average deviation of only 5\%.

($v$) The potential by Avrigeanu {\it et al.}\ \cite{Avr14} shows a stronger
energy dependence than the other potentials under study. This leads to a
better reproduction of the $(\alpha$,n) cross sections to the ground state at
higher energies, 
but significant underestimation for the isomeric ($\alpha$,n) cross sections
in particular at lower energies. For the ($\alpha$,$\gamma$) channel a similar
behavior is found with relatively poor $\chi^2$ and an increasing
underestimation towards lower energies; the E1 strength is here taken from the
hybrid model \cite{Gor98}. Overall, the Avrigeanu potential is only able to
reach $\chi^2 \approx 10$ 
although this potential provides the best description of the elastic
scattering data (see Sec.~\ref{sec:scat}).

Summarizing the above findings and applying the criterion $\chi^2 =
\chi^2_{\rm{min}} + 1$ (as discussed in detail in \cite{Mohr17}), the new
experimental data select the Atomki-V1 potential for the $\alpha$-OMP, but
different parameter combinations of nucleon-OMP, $\gamma$-ray strength
function, and level density reach the above criterion. In addition, the
$\alpha$-OMPs by Demetriou (Dem-1) and McFadden/Satchler remain very close to
the above criterion. The results from these $\alpha$-OMPs are also taken into
account in the following estimation of the ($\alpha$,$\gamma$) cross section
at astrophysically relevant low energies.

A more detailed study of sensitivities to the different ingredients of the
statistical model calculations is given in the Appendix
\ref{sec:app_sens}. From that study it can be concluded that the $\alpha$-OMPs
by Avrigeanu {\it et al.}\ in their earlier version \cite{Avr94} and by Nolte
{\it et al.}\ \cite{Nol87} are not able to reproduce the present experimental
data for $^{115}$In with reasonable $\chi^2$. This finding is not very
surprising as these potentials have been optimized mainly at higher energies.

For the prediction of $\alpha$-induced reaction cross sections at lower
energies and a discussion of the resulting uncertainties we select the
astrophysically relevant energy range, i.e.\ the classical Gamow window
energies for stellar temperatures of two and three billion degree ($T_9 = 2$
and 3) which are $E_0 = 6.46$ and 8.47 MeV in the center-of-mass system. Table
\ref{tab:extr} lists the extrapolated cross sections calculated using the
different $\alpha$-OMPs. At the lower energy $E_0 = 6.46$ MeV only the proton
and $\gamma$ channels are open; the neutron channel opens at about 7.2 MeV.
\begin{table*}
\caption{
\label{tab:extr} 
$\alpha$-induced cross sections at the energies corresponding to the Gamow
window at T$_9$ = 2 and 3 GK (in nb) calculated using different global
$\alpha$-nucleus potentials and either TALYS default settings (first line) or
the optimized parameter combinations as discussed in the text.
} 
\setlength{\extrarowheight}{0.1cm}
\begin{ruledtabular}
\begin{tabular}{ccccccccccc}
\parbox[t]{2.5cm}{\centering{potential}} &
\multicolumn{5}{c}{{\centering{E$_0$ = 6.46 MeV}}} &
\multicolumn{5}{c}{{\centering{E$_0$ = 8.47 MeV}}} \\
& $\sigma_{\rm{reac}}$ 
& ($\alpha$,$\gamma$)
& ($\alpha$,n)-gs
& ($\alpha$,n)-iso
& ($\alpha$,p) 
& $\sigma_{\rm{reac}}$ 
& ($\alpha$,$\gamma$)
& ($\alpha$,n)-gs
& ($\alpha$,n)-iso
& ($\alpha$,p) \\
\hline
Watanabe \cite{Wat58} \footnote{with TALYS default settings}      
& 15.5   & 15.4    & $-$    & $-$    & 0.06
& 11530  & 380     & 9140  & 2000   & 7    \\
Atomki-V1 \cite{Moh_ADNDT} 
& 1.3    & 1.3    & $-$    & $-$    & $\approx 1$ pb
& 3740   & 370    & 2930   & 440    & 3.6  \\
Demetriou-I \cite{Dem02}   
& 3.5    & 3.5    & $-$    & $-$    & $\approx 6$ pb
& 5100   & 480    & 3770   & 850    & 7.5  \\
McFadden \cite{Mcf}        
& 6.5    & 6.5    & $-$    & $-$    & 0.01
& 7730   & 570    & 6090   & 1060   & 7.1  \\
Avrigeanu \cite{Avr14}     
& 0.8    & 0.8     & $-$    & $-$    & $\approx 1$ pb
& 2520   & 370     & 1850   & 290    & 2.6  \\
\end{tabular} 
\end{ruledtabular}
\end{table*}

At the upper energy of 8.47 MeV, the total cross section $\sigma_{\rm{reac}}$
is relatively well-constrained around 400 nb from the Atomki-V1
potential with the smallest overall $\chi^2$, but also the Demetriou and
McFadden/Satchler potentials provide values around 500 nb. Even the TALYS
default calculation and the result from Avrigeanu lead to the same values
around 400 nb. Thus, at the upper energy of 8.47 MeV, the ($\alpha$,$\gamma$)
cross section is well constrained by the new experimental data with an
uncertainty of about 25\%.

At the lower energy of 6.46 MeV, the total cross section $\sigma_{\rm{reac}}$
is constrained between 1.3 and 6.5 nb, i.e.\ by a factor of about 5. The total
cross section is dominated by the ($\alpha$,$\gamma$) channel with very minor
contributions of the order of pico-barns for the ($\alpha$,p) channel. The
Atomki-V1 potential with its smallest $\chi^2$ predicts the small cross
section of 1.3 nb whereas the Demetriou and McFadden/Satchler potentials
predict higher cross sections of 3.5 nb and 6.5 nb. The TALYS default
calculation clearly overestimates the cross section with 15.5 nb, but
interestingly the Avrigeanu potential with its larger $\chi^2$ leads to a
prediction of 0.8 nb, only slightly lower than the best-fit prediction from
the Atomki-V1 potential. Combining the above findings, we recommend a cross
section of about $2^{+4}_{-1}$ nb at 6.46 MeV.

It has to be noted that the new experimental data are the prerequisite to
obtain the well-constrained cross sections of the ($\alpha$,$\gamma$) reaction
of 400 nb $\pm 25$\% at 8.47 MeV and $2^{+4}_{-1}$ nb at 6.46 MeV. Without the
$\chi^2$-assessment of the theoretical predictions, the range of predicted
cross sections is much larger. At the upper energy of 8.47 MeV, the
predictions from the different TALYS input parameters vary over more than two
orders of magnitude between 8 nb and 2.6 $\mu$b, and at the lower energy the
range is even larger with more than three orders of magnitude between 0.06 nb
and 100 nb. This wide range of predictions results not only from
uncertainties of the $\alpha$-OMP, but also from the chosen combination of the
nucleon-OMP, the $\gamma$-ray strength function, and the level density
parametrization. 

Finally, the relevance of preequilibrium reactions has to be discussed. In the
above calculations, the preequilibrium contribution was neglected. Including
preequilibrium leads to several findings and problems. An initial $\chi^2$
assessment including the preequilibrium contribution provided $\chi^2$ values
with a minimum of about 3.4, i.e.\ a factor of two worse than the above
results without preequilibrium. Consequently, the preequilibrium contribution
(as provided in TALYS) cannot be considered as reliable in the present case
and was therefore neglected in the above analysis.

The preequilibrium cross sections in TALYS are taken from an exciton
model. The results show a significant ($\alpha$,p) contribution at very low
energies which can be very close to the total reaction cross section
$\sigma_{\rm{reac}}$; e.g., at 6.46 MeV about 90\% of the total reaction cross
section go to the ($\alpha$,p) channel. As a consequence, the preequilibrium
calculations turned out to be numerically very delicate. Because the
($\alpha$,$\gamma$) cross section at low energies is essentially given by the
difference between the total reaction cross section $\sigma_{\rm{reac}}$ and
the ($\alpha$,p) cross section, and the latter is dominated by the
preequilibrium contribution, the ($\alpha$,$\gamma$) cross section becomes
extremely sensitive to the calculated ($\alpha$,p) preequilibrium cross
section. The default resolution of the TALYS energy grid had to be improved by
a factor of 10 to obtain numerically stable ($\alpha$,$\gamma$) cross sections
at low energies. This high sensitivity of the TALYS results on the treatment
of preequilibrium may also appear for other reactions. In the present case of
the $^{115}$In target, this sensitivity results from the relatively small
$Q$-value of the ($\alpha$,p) reaction which is typically strongly negative
for the recently studied even-even p-nuclei (e.g.,
\cite{Hal16,Sch16,Net15,Hal12,Rapp08,Ozk07,Gyu06,Som98}).

Further insight into the relevance of the preequilibrium contribution could be
obtained from low-energy ($\alpha$,p) data which are unfortunately not
accessible by activation. Also the reverse $^{118}$Sn(p,$\alpha$)$^{115}$In
reaction could be used to further constrain the parameter space; however, also
for the reverse reaction the data are very sparse. In the literature, only an
estimate of about 1 mb is provided at $E_p = 17$ MeV \cite{Chan73} which is
above the most relevant energy range of the present study.

\subsection{\label{sec:scat} Comparison to the elastic $\alpha$ scattering data}
In the course of the present investigation, also elastic scattering cross
sections of the $^{115}$In($\alpha,\alpha$)$^{115}$In reaction were studied at
energies E$_{Lab}$ = 16.15 MeV and E$_{Lab}$ = 19.50 MeV at Atomki
\cite{Kis16}. Complete angular distributions between 20$^{\circ}$ and
175$^{\circ}$ were measured at both energies in 1$^\circ$ ($20^\circ \leq \vartheta \leq 100^\circ$) and 2.5$^\circ$
($100^\circ \leq \vartheta \leq 175^\circ$) steps.
Total reaction cross sections $\sigma_{\rm{reac}}$ were derived from the fits
to the angular distributions, leading to $\sigma_{\rm{reac}} = 350.5 \pm 10.6$
mb ($777.0 \pm 23.5$ mb) at 16.15 MeV (19.50 MeV). 
In general, the results of the calculations using the various $\alpha$-nucleus
OMPs are in good agreement with the experimental results. Deviations do not
exceed about 15\%. The calculated elastic scattering angular distributions
are compared to the experimental scattering data in Fig.\ \ref{fig:an_di}, and
the total reaction cross sections $\sigma_{\rm{reac}}$ are listed in Table
\ref{tab:reac}. 
\begin{table}
\caption{\label{tab:reac} Total reaction cross sections $\sigma_{\rm{reac}}$
  (in mb) of predictions using different global parameterizations compared
  with $\sigma_{\rm{reac}}$ from the recently measured angular distributions
  \cite{Kis16}.
}
\setlength{\extrarowheight}{0.1cm}
\begin{ruledtabular}
\begin{tabular}{ccc}
\parbox[t]{2.0cm}{\centering{potential}} &
\parbox[t]{3.0cm}{\centering{E$_{Lab}$ = 16.15 MeV} \\ $\sigma_{\rm{reac}}$} &
\parbox[t]{3.0cm}{\centering{E$_{Lab}$ = 19.50 MeV} \\ $\sigma_{\rm{reac}}$} \\
\hline
experiment \cite{Kis16}    & $350.5 \pm 10.6$   & $777.0 \pm 23.5$  \\
Watanabe \cite{Wat58} \footnote{with TALYS default settings}      & 321.7              & 713.6  \\
Atomki-V1 \cite{Moh_ADNDT} & 404.1              & 815.2  \\
Demetriou-I \cite{Dem02}   & 335.9              & 741.5  \\
McFadden \cite{Mcf}        & 333.1              & 735.1  \\
Avrigeanu \cite{Avr14}     & 349.6              & 763.5  \\
\end{tabular} 
\end{ruledtabular}
\end{table}

As can be seen, at the lower energy E$_{Lab}$ = 16.15 MeV the angular
distribution is best reproduced by the potential of Avrigeanu
\cite{Avr14}. The potential by Demetriou (Dem-1) \cite{Dem02} and the early
potential by Watanabe \cite{Wat58} also provide very reasonable
descriptions. However, the experimental elastic cross sections are slightly
underestimated by the Atomki-V1 potential \cite{Moh_ADNDT} and slightly
overestimated by the McFadden/Satchler potential \cite{Mcf}.

The picture changes a bit for the higher energy E$_{Lab}$ = 19.50 MeV
angular distribution because of the increasing deviation from the Rutherford
cross section. Again, the measured data are very well reproduced by the
calculation using the potential of Avrigeanu \cite{Avr14}. From about
$45^\circ$ to $90^\circ$ the calculations with the potentials of McFadden
\cite{Mcf}, Demetriou \cite{Dem02}, and Watanabe \cite{Wat58} overestimate the
elastic cross sections, leading to an underestimation of the total reaction
cross section $\sigma_{\rm{reac}}$. Contrary, the calculation with the
Atomki-V1 \cite{Moh_ADNDT} slightly underestimates the elastic cross section
and thus overestimates $\sigma_{\rm{reac}}$. The somewhat larger deviations at
backward angles do not have major effects on $\sigma_{\rm{reac}}$.

Furthermore, the total reaction cross sections $\sigma_{\rm{reac}}$ from the
elastic scattering angular distributions (see Table \ref{tab:reac}) can be
included in the $\chi^2$ search for the best-fit parameters. These two
additional data points do not affect the conclusions of Sect.~\ref{sec:chi2}
although the $\chi^2$ of the Atomki-V1 potential increases from about 1.7 to
2.5; the $\chi^2$ values for the Demetriou (Dem-1) potential and the
McFadden/Satchler potential remain almost identical because of the better
reproduction of the experimental $\sigma_{\rm{reac}}$  in Table \ref{tab:reac}.

\subsection{Final remarks}
\label{sec:final}
The comparison between the experimental data of $\alpha$-induced reactions and
elastic scattering data to the calculations using various $\alpha$-nucleus
OMPs leads to several unexpected findings. In general, all $\alpha$-OMPs show
a quite reasonable agreement with the new experimental ($\alpha$,n) and
($\alpha$,$\gamma$) data. A similar result was found recently for
$\alpha$-induced reactions on the lighter target $^{64}$Zn \cite{Orn16,Mohr17}
whereas for heavier target nuclei often a significant overestimation of
reaction cross sections is found, in particular for the lowest
energies. However, despite the generally reasonable agreement, a strict
$\chi^2$-based assessment shows significant differences for the global
$\alpha$-OMPs under study.

The overall best-fit to the ($\alpha$,$X$) cross sections with $\chi^2 \approx
1.7$ is obtained from the Atomki-V1 potential. Slightly increased $\chi^2$ of
2.8 and 3.3 are found for the Demetriou (Dem-1) potential and the very simple
McFadden/Satchler potential which otherwise has a trend to overestimate
reaction data, in particular for heavy target nuclei towards lower energies. 
Contrary to the good description of the $\alpha$-induced reaction data, the
Atomki-V1 potential and the McFadden/Satchler potential show a relatively poor
reproduction of the elastic angular distributions. As these potentials,
Atomki-V1 and McFadden/Satchler, have been derived from the analysis of
elastic scattering data, one might expect better results for elastic
scattering and worse reproduction of reaction cross sections.

Contrary to the aforementioned Atomki-V1 and McFadden/Satchler potentials, the
potentials by Demetriou and by Avrigeanu have been derived from adjustments to
low-energy elastic scattering and reaction cross sections. The Demetriou
(Dem-1) potential provides a good description of the experimental reaction
data with $\chi^2 \approx 2.8$ whereas the Avrigeanu potential shows a
significantly increased overall $\chi^2 \approx 10$. However, in particular the
many-parameter potential of Avrigeanu provides an excellent reproduction of
the elastic scattering angular distributions.

Furthermore, the $\chi^2$ analysis is not able to provide final conclusions on
the other ingredients of the statistical model. The smallest $\chi^2$ for the
Atomki-V1, Dem-1, and McFadden/Satchler $\alpha$-OMPs are obtained with
different combinations of the nucleon OMP, $\gamma$-ray strength function, and
level density.

\section{Application of the best-fit parameters to $^{113}$In}
\label{sec:in113}
Recently, cross sections of $\alpha$-induced reactions for $^{113}$In have been
measured by  Yal\c c\i n {\it et al.}\ \cite{Yal09}, and the data have been
analyzed together with elastic scattering in \cite{Kis13}. For an excellent
reproduction of the experimental data, in both studies \cite{Yal09,Kis13} the
$\gamma$-ray strength had to be scaled down by at least 30\%.

\begin{figure}
\resizebox{0.75\columnwidth}{!}{\rotatebox{0}{\includegraphics[clip=]{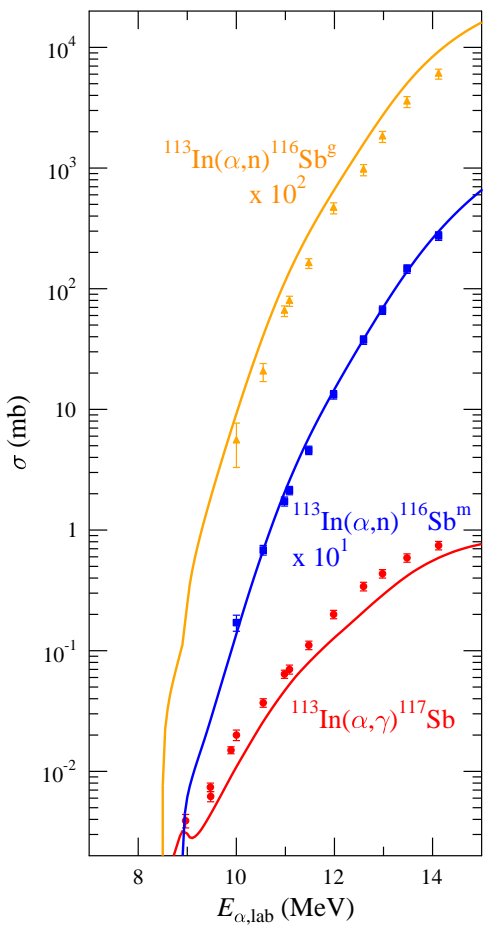}}}
\caption{
\label{fig:in113} 
(Color online) Application of the best-fit parameters from $^{115}$In to the
neighboring nucleus $^{113}$In: The experimental data \cite{Yal09} are
reasonably well reproduced without any further adjustment of parameters.
}
\end{figure}

Using the best-fit parameters of $^{115}$In from the present work, we find
reasonable agreement with the experimental data for $^{113}$In without any
scaling. The energy dependence of all cross sections is very well reproduced;
however, there is a slight overestimation of the ($\alpha$,n) cross section
and a slight underestimation of the ($\alpha$,$\gamma$) cross section. The
agreement for the isomeric contribution to the ($\alpha$,n) cross section is
excellent. The results for $^{113}$In are shown in Fig.~\ref{fig:in113}.

\section{\label{sec:sum} Summary and conclusions}

The $^{115}$In($\alpha,\gamma$)$^{119}$Sb and 
$^{115}$In($\alpha$,n)$^{118}$Sb$^m$ reaction cross sections were measured
between $E_\mathrm{c.m.}$ = 8.83 MeV - 15.58 MeV, and the
$^{115}$In($\alpha$,n)$^{118}$Sb$^{g}$ reaction was studied between
$E_\mathrm{c.m.}$ = 11.10 MeV - 15.58 MeV. Experimental results were compared
with Hauser-Feshbach statistical model calculations aiming the evaluation of
the global $\alpha$-nucleus optical potentials used in the weak r- and
$\gamma$ process studies.  

We found reasonable agreement between the experimental ($\alpha$,n) data and
the predictions using most global $\alpha$-nucleus optical potentials. The
best agreement was obtained for the Atomki-V1, the Demetriou (Dem-1), and the
McFadden/Satchler potentials. This fact clearly indicates that the alpha
transmission coefficient is reasonably well predicted for $^{115}$In by most
global potentials in the investigated energy interval. The experimental
($\alpha$,$\gamma$) cross sections are far underestimated by the TALYS default
parameters, but by selection of a different $\gamma$-ray strength
function also the ($\alpha$,$\gamma$) data could be well reproduced. The
extrapolation of the ($\alpha$,$\gamma$) cross section to the astrophyiscally
relevant low-energy region is nicely constrained by the new experimental data,
leading to a reduction of the uncertainty of about two orders of magnitude
when compared to the range of theoretical predictions within TALYS.

Surprisingly, the potential with the best description of the experimental
($\alpha$,$X$) reaction cross sections shows the largest deviations for the
elastic scattering angular distributions. Further efforts are needed to
establish an improved global $\alpha$-nucleus potential which simultaneously
describes elastic scattering and reaction data.

\section*{Acknowledgments}
This work was supported by NKFIH (K108459, K120666).

\appendix

\section{Further sensitivity studies}
\label{sec:app_sens}
As already mentioned in Sec.~\ref{sec:chi2}, it is impossible to show the
results of all 6720 combinations of input parameters for the statistical model
which were tested in the present study. In this Appendix we provide additional
information on the sensitivity of the cross sections for the different exit
channels, i.e., the ($\alpha$,$\gamma$) capture cross section and the
($\alpha$,n) cross sections feeding the ground state and the isomer in
$^{118}$Sb.

A simple qualitative discussion of the sensitivities of the calculated cross
sections was already provided at the beginning of Sec.~\ref{sec:chi2} which
was based on the definition of the reaction cross section in the statistical
model in Eq.~(\ref{eq:HF}) and the properties of the transmission
coefficients $T_i$ into the different exit channels. In the following we use
the best-fit calculation from Sec.~\ref{sec:chi2} (see also Table
\ref{tab:chi2}) as a reference and vary each ingredient of the statistical
model separately. 

\begin{figure}
\resizebox{0.75\columnwidth}{!}{\rotatebox{0}{\includegraphics[clip=]{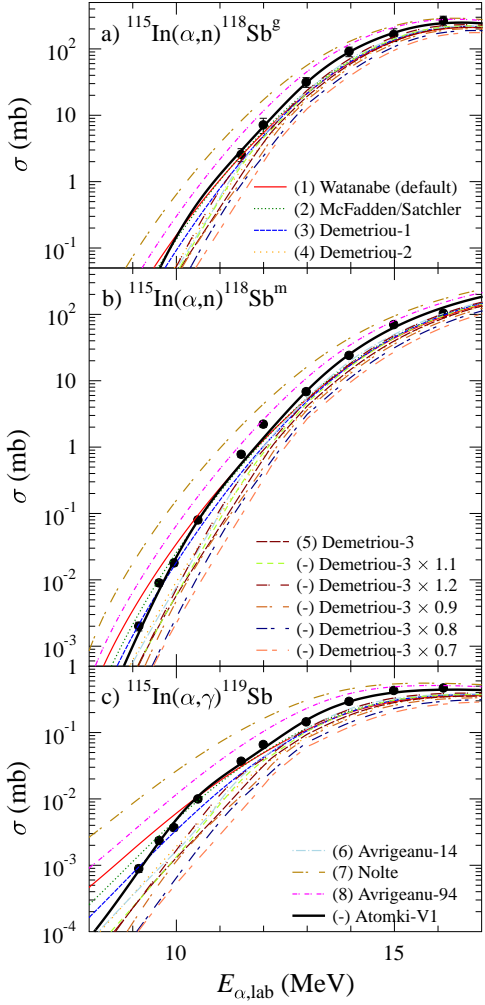}}}
\caption{
\label{fig:sens_aomp} 
(Color online) 
Sensitivity of the calculated cross sections to the chosen $\alpha$-OMP. All
other ingredients were kept fixed at the best-fit results in Table
\ref{tab:chi2}: 
a) $^{115}$In($\alpha$,n)$^{118}$Sb$^{g}$, 
b) $^{115}$In($\alpha$,n)$^{118}$Sb$^{m}$, 
c) $^{115}$In($\alpha$,$\gamma$)$^{119}$Sb. The numbers in the legend refer to
the TALYS numbering of $\alpha$-OMPs. The full black line shows the best-fit
from Table \ref{tab:chi2}.
}
\end{figure}

It is obvious from Eq.~(\ref{eq:HF}) that the transmission $T_{\alpha,0}$
affects all calculated cross sections. Thus, a significant sensitivity to
the $\alpha$-OMP is expected which is confirmed by the results in
Fig.~\ref{fig:sens_aomp}. At higher energies above 15 MeV the differences
between the various $\alpha$-OMPs become smaller whereas at low energies the
calculated cross sections vary over more than one order of magnitude. The
potential of Nolte {\it et al.}\ \cite{Nol87} and the earlier Avrigeanu
potential \cite{Avr94} overestimate the experimental data at low energies,
leading to huge $\chi^2$ for these potentials. The Demetriou potentials have a
trend to underestimate the experimental data; this trend becomes more
pronounced if the Demetriou potential is scaled down by factors smaller than
unity.

\begin{figure}
\resizebox{0.75\columnwidth}{!}{\rotatebox{0}{\includegraphics[clip=]{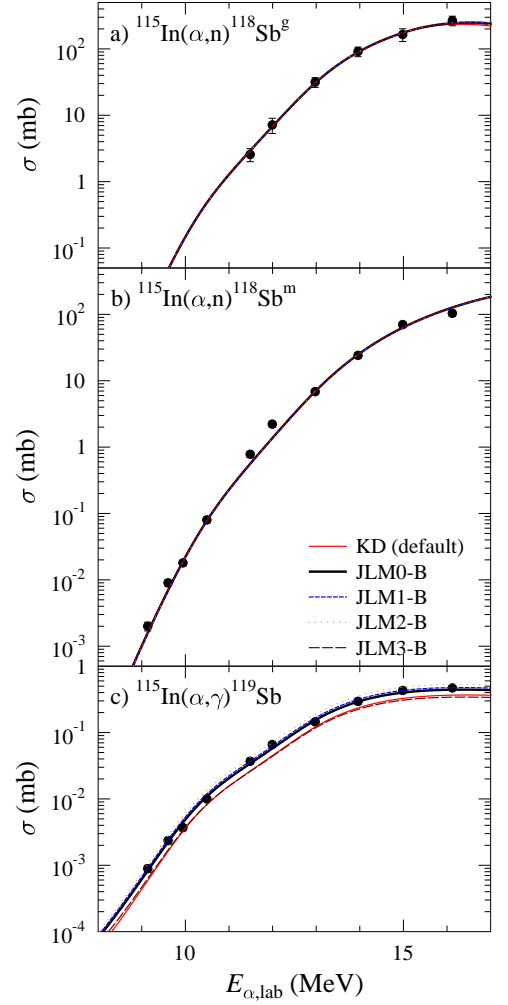}}}
\caption{
\label{fig:sens_nomp} 
(Color online) 
Same as Fig.~\ref{fig:sens_aomp} for the sensitivity to the nucleon OMPs.
}
\end{figure}

The sensitivity of the calculated cross sections to the other ingredients of
the statistical model is relatively weak. This holds in particular for the
nucleon OMP, as can be seen from Fig.~\ref{fig:sens_nomp}. As $T_n$ in
Eq.~(\ref{eq:HF}) is much larger than the other $T_i$, the branching $b_n$ to
the neutron channel becomes close to unity (independent of the absolute value
of $T_n$; see upper parts of Fig.~\ref{fig:sens_nomp}). The
$(\alpha$,$\gamma$) cross section is slightly influenced because the
($\alpha$,$\gamma$) cross section scales approximately with $T_{\alpha,0}
T_\gamma / T_n$ above the neutron threshold (lower part of
Fig.~\ref{fig:sens_nomp}).

\begin{figure}
\resizebox{0.75\columnwidth}{!}{\rotatebox{0}{\includegraphics[clip=]{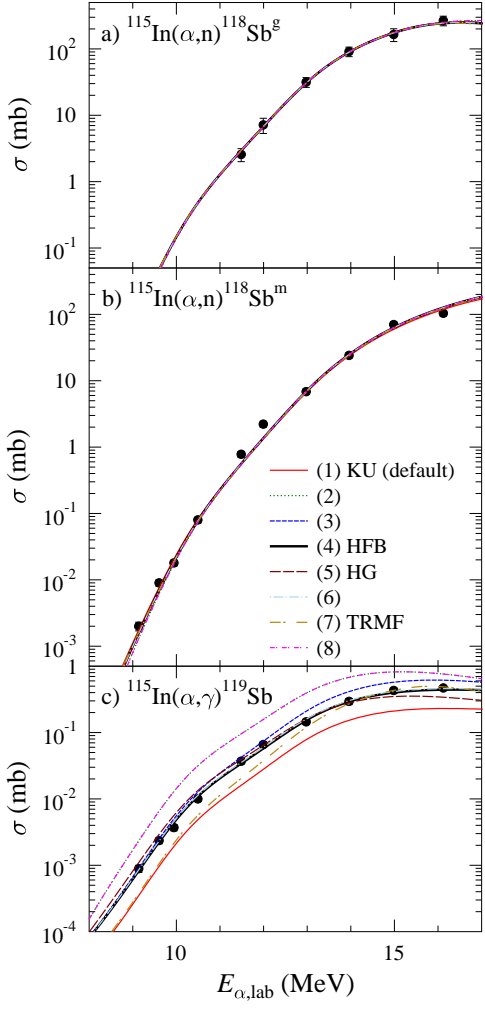}}}
\caption{
\label{fig:sens_str} 
(Color online) 
Same as Fig.~\ref{fig:sens_aomp} for the sensitivity to the $\gamma$-ray
strength functions. For references to the parameters beyond Table
\ref{tab:chi2}, see the TALYS manual \cite{TALYS}.
}
\end{figure}

As $T_\gamma$ is much smaller than $T_n$ above the neutron threshold, a
variation of the $\gamma$-ray strength does practically not affect the
cross section in the dominating neutron channel. However, the
($\alpha$,$\gamma$) cross section scales with $T_\gamma$. This is illustrated
in Fig.~\ref{fig:sens_str} which shows a variation of about a factor of $2-3$
around the best-fit calculation.

\begin{figure}
\resizebox{0.75\columnwidth}{!}{\rotatebox{0}{\includegraphics[clip=]{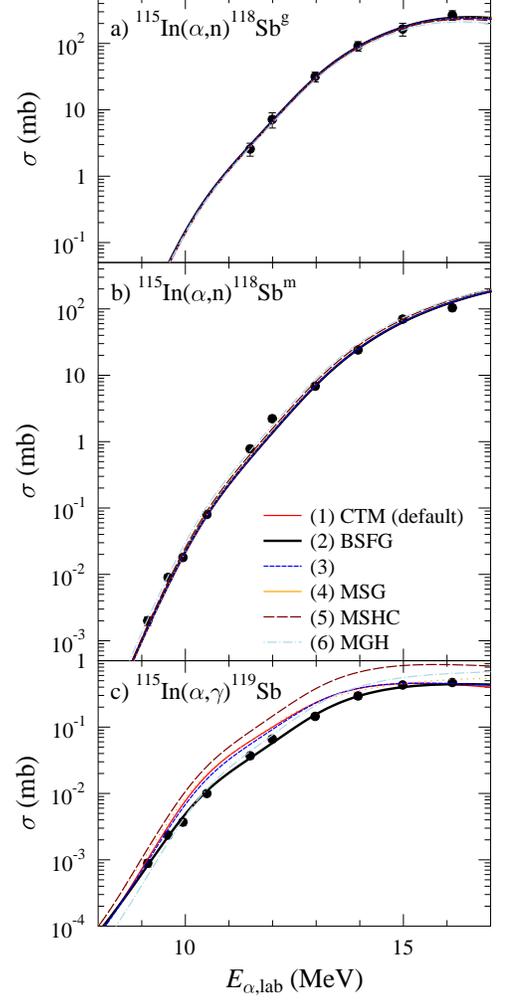}}}
\caption{
\label{fig:sens_ld} 
(Color online) 
Same as Fig.~\ref{fig:sens_aomp} for the sensitivity to the level densities.
For references to the parameters beyond Table \ref{tab:chi2}, see the TALYS
manual \cite{TALYS}.
}
\end{figure}

The role of the level densities is slightly different from the other above
ingredients because the chosen level density affects the calculated cross
section indirectly. Except $T_{\alpha,0}$, all $T_i$ in Eq.~(\ref{eq:HF})
result from the sum over all energetically allowed final states. In practice,
low-lying levels are taken into account explicitly, whereas the contributions
from higher-lying levels are taken into account by a theoretical level
density. Because of the negative $Q$-value of the ($\alpha$,n) reaction, in
the energy range under study practically all relevant levels are taken into
account explicitly, and thus the role of the level density is negligible in
the ($\alpha$,n) channel. The ($\alpha$,$\gamma$) cross section varies by
about a factor of two for the given 6 options for the level density in TALYS
(see Fig.~\ref{fig:sens_ld}).

Combining the above sensitivities, it can be concluded that a simultaneous
measurement of many reaction channels is essential to constrain the parameters
of the statistical model. Otherwise, different shortcomings of the chosen
parameters may (at least partly) compensate each other. This holds in
particular for the $(\alpha$,$\gamma$) channel with its sensitivity to
$T_{\alpha,0}$, $T_n$, and $T_\gamma$ and the implicit dependence on the level
density.

For completeness it can be noted that the various options for the nucleon-OMP,
the $\gamma$-ray strength, and the level density in combination with the
best-fit $\alpha$-OMP lead to a wide range in $\chi^2$ from about 1.7 for the
best-fit ($f_{\rm{dev}} = 1.12$) up to about 1500 ($f_{\rm{dev}} =
2.32$). This huge range for the total $\chi^2$ results mainly from the
($\alpha$,$\gamma$) channel where the worst calculation shows $\chi^2 \approx
4000$ and $f_{\rm{dev}} \approx 7$. The ($\alpha$,n) cross sections to the
$^{118}$Sb ground state and isomer are almost insensitive the the other
ingredients except the $\alpha$-OMP with a variation of $\chi^2$ of about a
factor of two. A variation of the $\gamma$-ray strength function (level
density) alone leads to a range of overall $\chi^2$ up to 105 (63) and
$f_{\rm{dev}} = 1.67$ (1.47). Interestingly, the overall best-fit is obtained
with an intermediate $\gamma$-ray strength and a low level density.
Alternatively, the data can be well described using a lower $\gamma$-ray
strength (e.g., the default KU strength, red line in Fig.~\ref{fig:sens_str})
with a higher level density (e.g., the MSHC level density, magenta dashed line
in Fig.~\ref{fig:sens_ld}) (for abbreviations see Table \ref{tab:chi2}).

As a final remark we point out that the sensitivities of the cross sections in
dependence of the input parameters in
Figs.~\ref{fig:sens_aomp}-\ref{fig:sens_ld} should be interpreted as empirical
sensitivities. These variations result from the range of predictions of widely
used global parametrizations. Contrary to the empirical sensitivities of the
present study, mathematical sensitivities have been studied e.g.\ in
\cite{rausensi} and are provided at \cite{non}. These mathematical
sensitivities, as defined in Eq.~(1) of \cite{rausensi}, have the advantage of
being well-defined, but do not take into account the reasonable physical range
of the input parameters. In that sense both approaches, the empirical
sensitivities of the present study and the mathematical sensitivities of
\cite{rausensi}, should be considered as complemetary.

\end{document}